\documentclass[aip,jcp,preprint,groupedaddress]{revtex4-1}%
\usepackage{graphicx}
\usepackage{tabularx}
\usepackage{dcolumn}
\usepackage{longtable}
\usepackage{tensor}
\usepackage{placeins}
\usepackage{bm}
\usepackage{amsmath}
\usepackage{amsfonts}
\usepackage{amssymb}
\usepackage{float}%
\providecommand{\U}[1]{\protect\rule{.1in}{.1in}}
\begin{document}
\title{Single-Hessian thawed Gaussian approximation}
\author{Tomislav Begu\v{s}i\'{c}}
\email{tomislav.begusic@epfl.ch}
\author{Manuel Cordova}
\author{Ji\v{r}\'i Van\'i\v{c}ek}
\email{jiri.vanicek@epfl.ch}
\affiliation{Laboratory of Theoretical Physical Chemistry, Institut des Sciences et
Ing\'enierie Chimiques, Ecole Polytechnique F\'ed\'erale de Lausanne (EPFL),
CH-1015, Lausanne, Switzerland}
\date{\today}

\begin{abstract}
To alleviate the computational cost associated with on-the-fly \textit{ab
initio} semiclassical calculations of molecular spectra, we propose the
single-Hessian thawed Gaussian approximation, in which the Hessian of the
potential energy at all points along an anharmonic classical trajectory is
approximated by a constant matrix. The spectra obtained with this
approximation are compared with the exact quantum spectra of a one-dimensional
Morse potential and with the experimental spectra of ammonia and
quinquethiophene. In all cases, the single-Hessian version performs almost as
well as the much more expensive on-the-fly \textit{ab initio} thawed Gaussian
approximation and significantly better than the global harmonic schemes.
Remarkably, unlike the thawed Gaussian approximation, the proposed method
conserves energy exactly, despite the time dependence of the corresponding
effective Hamiltonian, and, in addition, can be mapped to a
higher-dimensional time-independent classical Hamiltonian system. We also
provide a detailed comparison with several related approximations used for
accelerating prefactor calculations in semiclassical simulations.

\end{abstract}
\maketitle

\graphicspath{{"d:/Group Vanicek/Desktop/SingleHessian/figures/"}
{./figures/}{C:/Users/Jiri/Dropbox/Papers/Chemistry_papers/2019/SingleHessian/figures/}}

\section{\label{sec:intro}Introduction}

Simulation of vibrationally resolved electronic spectra of large polyatomic
molecules is a challenge for computational chemistry. The exact calculation is
impossible for most but smallest molecular systems due to the exponentially
scaling cost of computing the full potential energy surfaces of the electronic
states involved in the transition. In the well-known time-independent
formalism, the intensities of the individual vibronic transitions are
determined by the Franck--Condon factors, i.e., the squares of overlaps
between the vibrational eigenstates of the two electronic states, while the
frequencies of transitions are given by the differences of the corresponding
vibrational eigenvalues. A popular method for computing vibronic spectra
constructs global harmonic models of the two potential energy
surfaces.\cite{Santoro_Barone:2007, Santoro_Barone:2008,Barone_Santoro:2009}
Then, the vibrational functions, as well as their overlaps, are given
analytically. Anharmonic corrections can be included
perturbatively\cite{Bonness_Luis:2006,Luis_Kirtman:2004,Yang_Lin:2012,
Egidi_Barone:2017} or
variationally.\cite{Luis_Christiansen:2006,Bowman_Meyer:2008,Koziol_Krylov:2009,Meier_Rauhut:2015}
In smaller systems, it is feasible to apply anharmonic corrections to both
eigenstates and eigenvalues, which affects both positions and intensities of
vibronic transitions.\cite{Mok_Dyke:2000,Luis_Kirtman:2004,Koziol_Krylov:2009}
In larger systems, however, this is computationally challenging and the
anharmonic corrections are almost exclusively included only through the
frequencies, without affecting the Franck--Condon
factors.\cite{Egidi_Barone:2014, Egidi_Barone:2017,Biczysko_Wierzejewska:2018}

Time-dependent approaches, based on computing the dipole time correlation
function,\cite{Heller:1981a,book_Mukamel:1999,book_Tannor:2007} have also been
developed at different levels of accuracy, ranging from global harmonic
models\cite{Baiardi_Barone:2013} to exact quantum dynamics
methods\cite{Meyer_Worth:2009} on anharmonic potential energy surfaces. The
time-dependent formalism allows for an on-the-fly implementation, where the
potential data are evaluated only when needed, and therefore provides an
easier route to including anharmonicity. We focus our attention on the thawed
Gaussian approximation,\cite{Heller:1975,Grossmann:2006} which, as several
other
semiclassical\cite{Tatchen_Pollak:2009,Ceotto_Atahan:2009,Ceotto_Atahan:2009a,Wong_Roy:2011,Gabas_Ceotto:2017,Gabas_Ceotto:2018}
and quantum\cite{Ben-Nun_Martinez:2000,
Saita_Shalashilin:2012,Richings_Lasorne:2015} dynamics methods, has been
implemented in an on-the-fly fashion and combined with an \emph{ab initio}
evaluation of the potential.\cite{Wehrle_Vanicek:2014,Wehrle_Vanicek:2015} The
method assumes validity of the Born--Oppenheimer approximation and propagates
a Gaussian wavepacket in a locally harmonic potential constructed about the
current center of the wavepacket at each time step. This rather simple
propagation scheme, proposed by Heller as the first step beyond the global
harmonic approximation in the hierarchy of time-dependent methods,
was shown to work well for low or medium resolution electronic
spectra, where only short-time propagation of the wavepacket is
needed.\cite{Wehrle_Vanicek:2014,Wehrle_Vanicek:2015,Patoz_Vanicek:2018} To
further reduce the computational cost of on-the-fly \emph{ab initio}
calculations, one can employ a Hessian interpolation scheme, in which the
Hessians are evaluated only every several steps and interpolated in
between.\cite{Wehrle_Vanicek:2014}

Here, we propose a new approach, which still uses a fully anharmonic classical
trajectory to guide the Gaussian wavepacket but only a single Hessian to
propagate the width. Hence, this \textquotedblleft single-Hessian thawed
Gaussian approximation\textquotedblright\ further reduces the cost of spectra
calculations to that of a single classical trajectory. The method is validated
on a Morse potential as well as on full-dimensional on-the-fly \emph{ab
initio} simulations of the absorption spectrum of ammonia and emission
spectrum of quinquethiophene. The single-Hessian method performs better than
the global harmonic approaches and in some cases even better than the standard
thawed Gaussian approximation. Although the effective Hamiltonian
associated to the single-Hessian thawed Gaussian approximation is
time-dependent, we demonstrate---both analytically and numerically---that the
energy is conserved. Finally, we explore the relation between this
single-Hessian approach and similar well-known approximations to the prefactor
in the semiclassical Herman--Kluk initial value representation.

\section{\label{sec:theory}Theory}

\subsection{\label{subsec:spectra}Time-dependent approach to vibrationally
resolved electronic spectroscopy}

Let $|\psi(t)\rangle$ be a wavepacket
\begin{equation}
|\psi(t)\rangle=e^{-i\hat{H}t/\hbar}|\psi_{i}\rangle, \label{eq:psi_t}%
\end{equation}
propagated with a time-independent Hamiltonian
\begin{equation}
\hat{H} = H(\hat{q}, \hat{p}) = \frac{1}{2} \hat{p}^{T} \cdot m^{-1} \cdot
\hat{p} + V(\hat{q}),
\end{equation}
where $|\psi_{i}\rangle$ represents the initial state. Within the electric
dipole approximation, first-order perturbation theory, and assuming the Condon
approximation, vibrationally resolved electronic spectra can be computed from
the wavepacket autocorrelation function
\begin{equation}
C(t)=\langle\psi_{i}|\psi(t)\rangle. \label{eq:C_t}%
\end{equation}
The type of spectroscopy determines the choice of $\psi_{i}$ and $H$. If
$\psi_{i}$ is a vibrational eigenstate of the ground electronic state $1$ and
$H$ the excited-state vibrational Hamiltonian, the rotationally averaged
absorption cross-section is evaluated as the Fourier
transform\cite{Heller:1981a,Lami_Santoro:2004,book_Tannor:2007}
\begin{equation}
\sigma_{\text{abs}}(\omega)=\frac{4\pi\omega}{3\hbar c}|\vec{\mu}_{21}%
|^{2}\text{Re}\int_{0}^{\infty}C(t)e^{i(\omega+E_{1,i}/\hbar)t}dt,
\label{eq:sigma_abs}%
\end{equation}
where $E_{1,i}$ is the energy of state $\psi_{i}$ before photon absorption and
$\vec{\mu}_{21}$ the transition dipole moment between the ground and excited
electronic states evaluated at the ground-state equilibrium geometry. The
emission spectrum, measured as the emission rate per unit frequency, is
obtained by taking the $\psi_{i}$ to be the vibrational eigenstate of the
excited electronic state $2$ and $H$ the ground-state vibrational
Hamiltonian:\cite{Lami_Santoro:2004,Niu_Shuai:2010}
\begin{equation}
\sigma_{\text{em}}(\omega)=\frac{4\omega^{3}}{3\pi\hbar c^{3}}|\vec{\mu}%
_{21}|^{2}\text{Re}\int_{0}^{\infty}C(t)^{\ast}e^{i(\omega-E_{2,i}/\hbar)t}dt,
\label{eq:sigma_em}%
\end{equation}
where $E_{2,i}$ is the energy of state $\psi_{i}$ before photon emission.
Spectra defined in Eqs. (\ref{eq:sigma_abs}) and (\ref{eq:sigma_em}) are
positive at all frequencies, which can be shown by inserting a resolution of
identity in the expression (\ref{eq:C_t}) for the autocorrelation function to
derive the time-independent expression; e.g., for the absorption spectrum, one
obtains\cite{book_Tannor:2007}
\begin{equation}
\sigma_{\text{abs}}(\omega)=\frac{4\pi^{2}\omega}{3\hbar c}|\vec{\mu}%
_{21}|^{2}\sum_{n}|\langle n|\psi_{i}\rangle|^{2}\delta(\omega-\frac
{E_{2,n}-E_{1,i}}{\hbar}), \label{eq:sigma_abs_ti}%
\end{equation}
where $|n\rangle$ are the eigenstates of the excited-state vibrational
Hamiltonian with energies $E_{2,n}$. Equations~(\ref{eq:sigma_abs}) and
(\ref{eq:sigma_abs_ti}) are equivalent for any time-independent Hamiltonian.
However, if the true time-independent Hamiltonian is approximated by an
effective time-dependent one, for example, through the local harmonic or cubic
approximations, negative spectral features may
arise.\cite{Wehrle_Vanicek:2015}

\subsection{\label{subsec:tga}Thawed Gaussian approximation}

Evaluation of the autocorrelation function (\ref{eq:C_t}) requires propagating
the vibrational wavepacket; among many quantum and semiclassical methods, one
of the simplest is the thawed Gaussian approximation.\cite{Heller:1975} A
thawed Gaussian wavepacket is described by its time-dependent position $q_{t}%
$, momentum $p_{t}$, complex symmetric matrix $A_{t}$, and a complex number
$\gamma_{t}$:
\begin{equation}
\psi(q,t)=N_{0}\exp\left\{  \frac{i}{\hbar}\left[  \frac{1}{2}(q-q_{t}%
)^{T}\cdot A_{t}\cdot(q-q_{t})+p_{t}^{T}\cdot(q-q_{t})+\gamma_{t}\right]
\right\}  , \label{eq:GWP}%
\end{equation}
where $N_{0}=[\det(\text{Im}A_{0}/\pi\hbar)]^{1/4}$ is a normalization
constant. Classical parameters $q_{t}$ and $p_{t}$ are the expectation values
of the position and momentum; the imaginary part of matrix $A_{t}$ controls
the width of the wavepacket, while its real part introduces a spatial chirp;
the real part of $\gamma_{t}$ is a time-dependent phase factor, while its
imaginary part ensures normalization at all times. The Gaussian form
(\ref{eq:GWP}) is exactly preserved under evolution in a harmonic potential,
even a time-dependent one. In the thawed Gaussian approximation, the
wavepacket (\ref{eq:GWP}) is propagated with an effective Hamiltonian $\hat
{H}_{\text{eff}}(t)=\hat{T}+\hat{V}_{\text{LHA}}(t)$ given by the sum of the
kinetic energy $T$ and the time-dependent local harmonic approximation
$V_{\text{LHA}}$ of the true potential $V$ about $q_{t}$:
\begin{equation}
V_{\text{LHA}}(q,t)=V(q_{t})+V^{\prime}(q_{t})^{T}\cdot(q-q_{t})+\frac{1}%
{2}(q-q_{t})^{T}\cdot V^{\prime\prime}(q_{t})\cdot(q-q_{t}), \label{eq:LHA}%
\end{equation}
with $V^{\prime}(q_{t})$ representing the gradient and $V^{\prime\prime}%
(q_{t})$ the $D\times D$ Hessian matrix of the potential evaluated at the
center of the wavepacket $q_{t}$. Inserting the Gaussian ansatz (\ref{eq:GWP})
and effective potential (\ref{eq:LHA}) into the time-dependent Schr\"{o}dinger
equation gives the following equations of motion for the wavepacket
parameters:\cite{Heller:1975}
\begin{align}
\dot{q}_{t}  &  =m^{-1}\cdot p_{t}\,,\label{eq:q_dot}\\
\dot{p}_{t}  &  =-V^{\prime}(q_{t})\,,\label{eq:p_dot}\\
\dot{A}_{t}  &  =-A_{t}\cdot m^{-1}\cdot A_{t}-V^{\prime\prime}(q_{t}%
)\,,\label{eq:A_dot}\\
\dot{\gamma}_{t}  &  =L_{t}+\frac{i\hbar}{2}\text{Tr}\left(  m^{-1}\cdot
A_{t}\right)  \,, \label{eq:gamma_dot}%
\end{align}
where $m$ is the mass matrix and $L_{t}$ the Lagrangian.

If the wavepacket remains localized, the effective locally harmonic potential
is a good approximation and the thawed Gaussian propagation is expected to be
rather accurate. The approximation accounts partially for anharmonicity by
propagating the wavepacket's center $(q_{t},p_{t})$ classically with the true,
anharmonic potential $V(q)$ [Eqs.~(\ref{eq:q_dot})--(\ref{eq:p_dot})
are Hamilton's equations of motion for $H(q,p)$] and by accounting for the
changes in its Hessian, which affect the semiclassical parameters $A_{t}$ and
$\gamma_{t}$. Another advantage of the thawed Gaussian approximation is its
efficiency: it requires propagating four time-dependent parameters, which
depend only on the local potential information.

Yet, there are also several drawbacks: First, the Gaussian ansatz
(\ref{eq:GWP}) cannot describe wavepacket splitting, tunneling, or
nonadiabatic effects. In very anharmonic systems, where the exact wavepacket
splits and delocalizes quickly, the thawed Gaussian wavepacket behaves
unphysically. Thus, the method is limited to short propagation
times and low-resolution electronic spectra. Second, because the effective
potential~(\ref{eq:LHA}) is, in general (i.e., for potentials beyond
quadratic), time-dependent, the thawed Gaussian approximation does not
conserve energy:\cite{Wehrle_Vanicek:2015,Rohrdanz_Cina:2006}
\begin{align}
\frac{dE}{dt}  &  =\frac{d}{dt}\langle\psi(t)|\hat{H}_{\text{eff}}%
(t)|\psi(t)\rangle\label{eq:de_dt_a}\\
&  =\langle\psi(t)|\frac{d}{dt}\hat{H}_{\text{eff}}(t)|\psi(t)\rangle
\label{eq:de_dt_b}\\
&  =\langle\psi(t)|\frac{d}{dt}\hat{V}_{\text{LHA}}(t)|\psi(t)\rangle
\label{eq:de_dt_c}\\
&  =\frac{1}{2}\langle\psi(t)|(\hat{q}-q_{t})^{T}\cdot B_{t}\cdot(\hat
{q}-q_{t})|\psi(t)\rangle\label{eq:de_dt_d}\\
&  =\frac{1}{2}\operatorname{Tr}(B_{t}\cdot\Sigma_{t}^{2}), \label{eq:de_dt_e}%
\end{align}
where $B_{t}:=p_{t}^{T}\cdot m^{-1}\cdot V^{\prime\prime\prime}(q_{t})$,
$V^{\prime\prime\prime}(q_{t})$ is a rank-3 tensor of third derivatives of the
potential with respect to position, and
\begin{align}
\Sigma_{t}^{2}  &  :=\langle\psi(t)|(\hat{q}-q_{t})\otimes(\hat{q}-q_{t}%
)^{T}|\psi(t)\rangle\\
&  =\int dq|\psi(q,t)|^{2}(q-q_{t})\otimes(q-q_{t})^{T}\\
&  =\bigg(\frac{2}{\hbar}\operatorname{Im}A_{t}\bigg)^{-1}%
\end{align}
is the position covariance matrix. Equation (\ref{eq:de_dt_b}) follows because
the thawed Gaussian solves exactly the Schr\"{o}dinger equation with
$H_{\text{eff}}$, while Eq.~(\ref{eq:de_dt_c}) relies on the time independence
of the kinetic energy operator. To derive Eq.~(\ref{eq:de_dt_d}), we used the
chain rule
\begin{equation}
\frac{d}{dt}=p_{t}^{T}\cdot m^{-1}\cdot\frac{d}{dq_{t}} \label{eq:chain_rule}%
\end{equation}
for the differentiation of the energy, gradient, and Hessian evaluated at
position $q_{t}$. As noted already in Section \ref{subsec:spectra}, the time
dependence of the effective Hamiltonian also leads to unphysical negative
intensities in the spectra.

\subsection{\label{subsec:interpolatedtga}Hessian interpolation}

To reduce the cost of \emph{ab initio} Hessian calculations, the on-the-fly
\emph{ab initio} thawed Gaussian approximation is readily combined with an
interpolation scheme, where the Hessians are computed only every few steps and
the intermediate Hessians are obtained from a second-order polynomial
interpolation. Typically, the Hessians need to be computed only every four to
eight time steps.\cite{Wehrle_Vanicek:2014, Begusic_Vanicek:2018a} Since the
Hessians are not needed for the propagation of the classical trajectory,
additional speed-up is achieved through parallel computation of the Hessians
after the full trajectory is known. Note that other Hessian approximations,
such as the Hessian update schemes\cite{Ceotto_Hase:2013,Zhuang_Ceotto:2013,
Ianconescu_Pollak:2013} and Gaussian process
regression\cite{Alborzpour_Habershon:2016,Laude_Richardson:2018} have been
developed in the context of \emph{ab initio} simulations. The considerable
cost of multiple Hessian evaluations has also inspired various semiclassical
approximations,\cite{DiLiberto_Ceotto:2016} including the
prefactor-free,\cite{Tatchen_Pollak:2009}
adiabatic,\cite{Guallar_Miller:1999,Guallar_Miller:2000}
harmonic,\cite{DiLiberto_Ceotto:2016} and \textquotedblleft poor
person's\textquotedblright\cite{Tatchen_Miller:2011} variations of the
Herman--Kluk propagator.

\subsection{\label{subsec:globalharmonic}Global harmonic approximation}

In computational chemistry, most calculations of vibrationally resolved
electronic spectra employ the global harmonic models, where the true potential
energy surface is approximated as%
\begin{equation}
V_{\text{HA}}(q)=V_{\text{eq}}+\frac{1}{2}(q-q_{\text{eq}})^{T}\cdot
k\cdot(q-q_{\text{eq}}). \label{eq:HA}%
\end{equation}
In Eq.~(\ref{eq:HA}), $V_{\text{eq}}$ is the potential energy and
$q_{\text{eq}}$ the position of the minimum of the harmonic potential with a
force constant matrix $k$. In practice, the global harmonic model is
constructed from \emph{ab initio} data evaluated at a single molecular
geometry, which makes such calculations feasible for rather large systems. The
thawed Gaussian wavepacket (\ref{eq:GWP}) is exact in the harmonic potential
(\ref{eq:HA}) and can be propagated analytically. Furthermore, because the
potential is time-independent, the energy is conserved exactly and the
corresponding spectra do not suffer from unphysical negative intensities.
However, the method neglects anharmonicity completely and, therefore, is less
accurate than the thawed Gaussian approximation.

\subsection{\label{subsec:shtga}Single-Hessian thawed Gaussian approximation}

Let us now consider using a single Hessian in the local harmonic
approximation~(\ref{eq:LHA}), e.g., by choosing a reference Hessian
$V_{\text{ref}}^{\prime\prime}(q_{\text{ref}})$ and approximating the
potential at each point in time as
\begin{equation}
V_{\text{SH}}(q,t)=V(q_{t})+V^{\prime}(q_{t})^{T}\cdot(q-q_{t})+\frac{1}%
{2}(q-q_{t})^{T}\cdot V_{\text{ref}}^{\prime\prime}(q_{\text{ref}}%
)\cdot(q-q_{t}). \label{eq:SHLHA}%
\end{equation}
The single-Hessian thawed Gaussian approximation, which propagates the
wavepacket~(\ref{eq:GWP}) in the effective potential~(\ref{eq:SHLHA}), is,
obviously, even more efficient than the original thawed Gaussian
approximation; the single-Hessian analogue requires only one Hessian to be
evaluated for the whole propagation, i.e., its cost is almost the same as
running a single classical trajectory. Because the effective
potential~(\ref{eq:SHLHA}) is Hermitian, the single-Hessian method conserves
the norm of the wavefunction. As for the accuracy, the
approximation~(\ref{eq:SHLHA}) of the potential still includes anharmonicity
partially through the first two terms and thus is more accurate than the
global harmonic approximation, but is clearly worse than the local harmonic
approximation~(\ref{eq:LHA}) (see Fig.~\ref{fig:HierarchyScheme}). Yet, the
single-Hessian approach also results in several improvements related to
spectra calculations:

\begin{figure}
\includegraphics[width=0.45\textwidth]{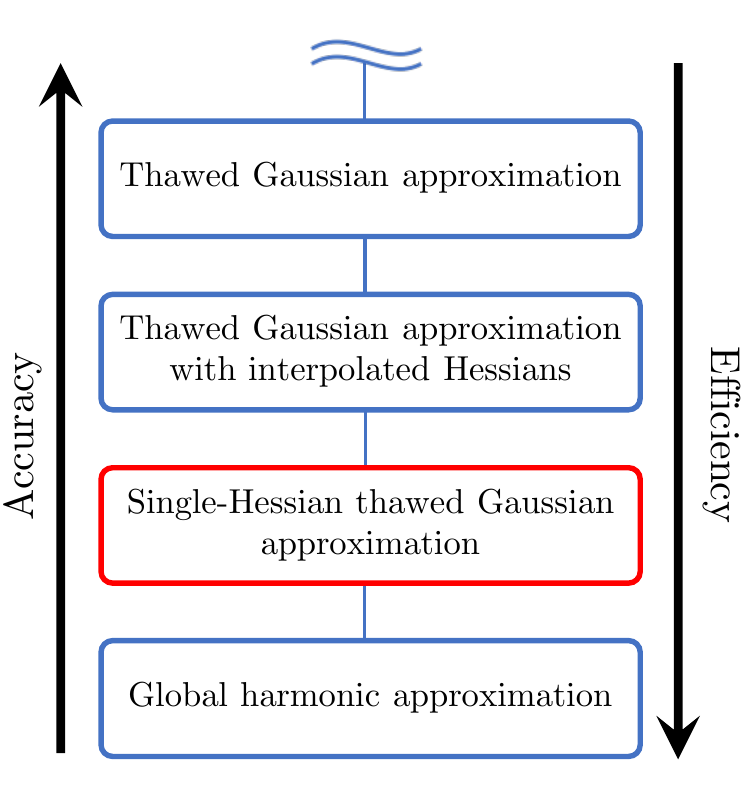}
\caption{\label{fig:HierarchyScheme}Hierarchy of several semiclassical wavepacket methods for simulating
vibrationally resolved electronic spectra. Various methods, which are more accurate, but also more
expensive than the thawed Gaussian approximation, are beyond the scope of this work, although they may describe
even high-resolution spectra.}
\end{figure}

First, the propagation of matrix $A_{t}$ is now determined exclusively by the
reference Hessian and is decoupled from the classical dynamics of $q_{t}$ and
$p_{t}$. Therefore, the wavepacket does not spread or contract unphysically in
an attempt to describe wavepacket splitting, but rather stays compact at all
times, similarly to a squeezed state in a globally harmonic
potential. We show in several numerical examples that this feature is
preferred in more anharmonic potentials.

Second, the single-Hessian thawed Gaussian approximation conserves energy
exactly:
\begin{align}
\frac{dE}{dt}  &  =\langle\psi(t)|\frac{d}{dt}\hat{V}_{\text{SH}}%
(t)|\psi(t)\rangle\label{eq:de_dt_sh_a}\\
&  =\langle\psi(t)|b_{t}^{T}\cdot(\hat{q}-q_{t})|\psi(t)\rangle
\label{eq:de_dt_sh_b}\\
&  =b_{t}^{T}\cdot\langle\psi(t)|\hat{q}-q_{t}|\psi(t)\rangle
\label{eq:de_dt_sh_c}\\
&  =0, \label{eq:de_dt_sh_d}%
\end{align}
where $b_{t}^{T}:=p_{t}^{T}\cdot m^{-1}\cdot(V^{\prime\prime}(q_{t}%
)-V_{\text{ref}}^{\prime\prime}(q_{\text{ref}}))$. Above, we used the time
independence of the kinetic energy operator in Eq.~(\ref{eq:de_dt_sh_a}) and
the chain rule (\ref{eq:chain_rule}) to go from Eq.~(\ref{eq:de_dt_sh_a}) to
(\ref{eq:de_dt_sh_b}). The final result (\ref{eq:de_dt_sh_d}) follows from
Eq.~(\ref{eq:de_dt_sh_c}) because $q_{t}$ is the expectation value of the
position operator $\hat{q}$ in the state $\psi(t)$. Despite the energy
conservation, the effective Hamiltonian determined by the effective potential
of Eq.~(\ref{eq:SHLHA}) is still time-dependent---the energy is conserved only
because the Hamiltonian is nonlinear (i.e., it depends on the state $\psi$)
and its change applied to $\psi$ happens to be \textquotedblleft
orthogonal\textquotedblright\ to the state $\psi$ [Eqs.~(\ref{eq:de_dt_sh_a}%
)--(\ref{eq:de_dt_sh_d})]. Therefore, the conservation of energy does not
guarantee non-negative intensities in the spectrum. Yet, the hope is that the
negative spectral features in the single-Hessian approach will be less
pronounced than in the standard thawed Gaussian approximation.

The single-Hessian thawed Gaussian approximation may seem to be
only a special, constant-Hessian case of one of several approximations used
for accelerating semiclassical calculations based on the Herman-Kluk
propagator.\cite{Herman_Kluk:1984} In Appendix~\ref{sec:app_HK}, we,
therefore, compare the proposed method with the single-Hessian versions of the
Herman--Kluk, Johnson's,\cite{Gelabert_Miller:2000} frozen
Gaussian,\cite{Heller:1981} adiabatic Herman--Kluk,\cite{Guallar_Miller:1999}
and prefactor-free\cite{Tatchen_Pollak:2009} approximations and show that the
equivalence holds only for some of these methods and, moreover, only if the
thawed Gaussian becomes \textquotedblleft frozen,\textquotedblright\ which
requires a specific choice of the reference Hessian.

\subsection{\label{subsec:refhess}Reference Hessians}

\begin{figure}
\includegraphics[width=0.4\textwidth]{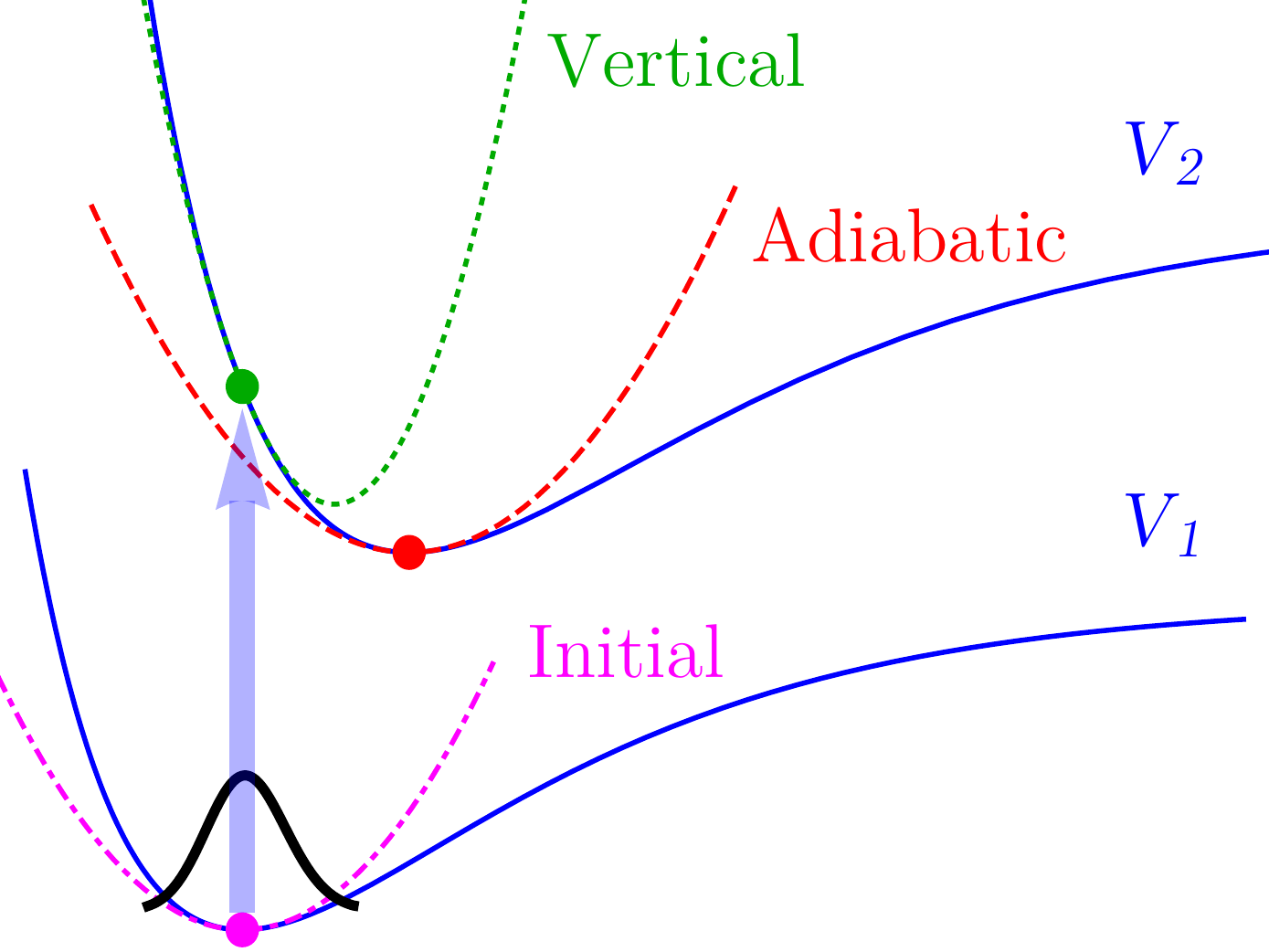}
\caption{\label{fig:VHAHInit_Scheme}Different choices of the reference Hessian.
The Hessian of the final-state surface is commonly evaluated at the Franck--Condon
(vertical Hessian, green dot) or minimum energy position (adiabatic Hessian, red dot), giving rise to the vertical
(green dotted curve) and adiabatic (red dashed curve) global harmonic models. Initial-state Hessian,
evaluated at the minimum of the initial-state surface (magenta dot) is needed for constructing the initial
wavefunction (black), given by the ground vibrational eigenstate of the harmonic fit (magenta dash-dotted curve)
to the initial-state surface. However, the initial-state Hessian can also serve as a crude approximation to the
final-state Hessian, resulting in the vertical gradient and adiabatic shift global harmonic models, or initial
single-Hessian thawed Gaussian approximation (see text).}
\end{figure}

Both global harmonic models and single-Hessian thawed Gaussian approximation
depend on the choice of the reference Hessian. Two well-known
special choices are the adiabatic Hessian---Hessian of the final electronic
potential energy surface evaluated at its minimum ($V_{\text{ref}%
}=V_{\text{final}}$, $q_{\text{ref}}=q_{\text{eq,\ final}}$), and the vertical
Hessian---Hessian of the final electronic surface evaluated at the
Franck--Condon point, i.e., the minimum of the initial electronic surface
($V_{\text{ref}}=V_{\text{final}}$, $q_{\text{ref}}=q_{\text{eq,\ init}}$);
see Fig.~\ref{fig:VHAHInit_Scheme}%
.\cite{AvilaFerrer_Santoro:2012,Egidi_Barone:2014} We refer to the
combinations of these two Hessian choices with the global harmonic approach as
the \emph{adiabatic harmonic} and \emph{vertical harmonic}
methods.\cite{Wehrle_Vanicek:2014,Patoz_Vanicek:2018,Begusic_Vanicek:2018} In
the literature, these global harmonic models are sometimes referred to as the
adiabatic and vertical
Hessian;\cite{AvilaFerrer_Santoro:2012,Egidi_Barone:2014,Baiardi_Barone:2013}
here, we use these names exclusively for the Hessians themselves to avoid the
confusion between the single-Hessian thawed Gaussian propagation and global
harmonic methods. The combinations of the single-Hessian approach with the
different reference Hessians will be referred to as the \emph{adiabatic
single-Hessian} and \emph{vertical single-Hessian} methods.

Finally, one can avoid computing any Hessian of the final
electronic surface by using as reference the initial-state Hessian---Hessian
of the initial electronic surface at its minimum ($V_{\text{ref}%
}=V_{\text{init}}$, $q_{\text{ref}}=q_{\text{eq,\ init}}$, see
Fig.~\ref{fig:VHAHInit_Scheme}), which is commonly needed already for
constructing the initial wavepacket. In the context of global harmonic
methods, there are two natural possibilities of constructing a final-state
harmonic potential using the initial-state Hessian: one can either compute the
potential energy and gradient of the final-state potential energy surface at
the initial geometry, which results in the \emph{vertical gradient} model, or
optimize the geometry in the final electronic state, which gives the
\emph{adiabatic shift}
model.\cite{AvilaFerrer_Santoro:2012,Cerezo_Santoro:2013,Egidi_Barone:2014}
Both the vertical gradient and adiabatic shift models are examples of
displaced harmonic systems, and thus ignore mode distortion and mixing (the
Duschinsky effect) between the two electronic states. In the results section,
we discuss only the adiabatic shift model and, for consistency with the other
methods discussed in this work, refer to it as the \emph{initial harmonic} model.

\subsection{\label{subsec:H_map}$PQ$-method and the Hamiltonian structure}

The Riccati equation (\ref{eq:A_dot}) can be solved with the \textquotedblleft%
$PQ$ method\textquotedblright, i.e., by introducing auxiliary complex $D\times
D$ matrices $Q_{t}$ and $P_{t}$ such that\cite{Heller:1976a}
\begin{align}
A_{t} &  =P_{t}\cdot Q_{t}^{-1},\label{eq:A_QP}\\
P_{t} &  =m\cdot\dot{Q}_{t}.\label{eq:P_t}%
\end{align}
Inverting Eq.~(\ref{eq:P_t}) and inserting Eq.~(\ref{eq:A_QP}) into
Eq.~(\ref{eq:A_dot}) yields the differential equations
\begin{align}
\dot{Q}_{t} &  =m^{-1}\cdot P_{t},\label{eq:Q_dot}\\
\dot{P}_{t} &  =-V^{\prime\prime}(q_{t})\cdot Q_{t},\label{eq:P_dot}%
\end{align}
which can be recognized as Hamilton's equations of motion%
\[
\dot{Q}_{t}=\frac{\partial H_{\text{sc}}}{\partial P^{\ast}}\text{ \ and
\ }\dot{P}_{t}=-\frac{\partial H_{\text{sc}}}{\partial Q^{\ast}}%
\]
of a \textquotedblleft semiclassical\textquotedblright\ time-dependent
Hamiltonian
\begin{equation}
H_{\text{sc}}(Q,P;q_{t})=\frac{1}{2}\operatorname*{Tr}[P^{\dagger}\cdot
m^{-1}\cdot P+Q^{\dagger}\cdot V^{\prime\prime}(q_{t})\cdot Q],\label{eq:H_sc}%
\end{equation}
where $q_{t}$ plays a role of an external, time-dependent parameter. Above,
$^{\ast}$ denotes a complex conjugate and $^{\dagger}$ the Hermitian
transpose, i.e., a complex conjugate and transpose of a matrix. Hamilton's
equations (\ref{eq:Q_dot})-(\ref{eq:P_dot}) solve Eq.~(\ref{eq:A_dot}) for
$\dot{A}_{t}$ for any choice of $Q_{0}$ and $P_{0}$ that satisfy
Eq.~(\ref{eq:A_QP}) at time zero.

In the single-Hessian approximation, the time-dependent Hessian is replaced
with the reference Hessian, and the semiclassical Hamiltonian (\ref{eq:H_sc})
becomes independent of $q_{t}$ and, therefore, also independent of time. As a
result, the quantum propagation using single-Hessian thawed Gaussian
approximation for $H(q,p)$ can be mapped to \emph{exact} classical propagation
with a separable Hamiltonian
\begin{equation}
H_{\text{map}}(q,p,Q,P)=H(q,p)+H_{\text{sc}}(Q,P). \label{eq:H_map}%
\end{equation}
Because of separability, both $H(q_{t},p_{t})$ and $H_{\text{sc}}(Q_{t}%
,P_{t})$ are independent of time. In Appendix~\ref{sec:app_H_map}, we show
that the energy $E(t)$ of the wavepacket (\ref{eq:GWP}) is equal to
$H_{\text{map}}(q_{t},p_{t},Q_{t},P_{t})$ for a specific choice of $Q_{0}$ and
$P_{0}$ (up to a factor equal to Hagedorn
parametrization\cite{Hagedorn:1980,Hagedorn:1998,book_Lubich:2008}), providing
an independent proof of energy conservation by the single-Hessian thawed
Gaussian approximation. Neither energy conservation nor mapping to a classical
Hamiltonian system holds for the original thawed Gaussian approximation due to
the dependence of the Hessian on $q_{t}$; in that case, Hamilton's equation
for $p_{t}$ derived from $H_{\text{map}}$ has an additional term compared to
Eq.~(\ref{eq:p_dot}). Yet, a similar mapping, yielding a nonseparable
Hamiltonian, does
exist\cite{Kramer_Saraceno:1981,Arickx_VanLeuven:1986,Faou_Lubich:2006} if one
applies the time-dependent variational
principle\cite{Heller:1976,Coalson_Karplus:1990} instead of the local harmonic
approximation (\ref{eq:LHA}) to the quantum propagation of the Gaussian
wavepacket (\ref{eq:GWP}).

\section{\label{sec:compdet}Computational details}

\subsection{\label{subsec:morse_comp}Morse potential}

To investigate the single-Hessian thawed Gaussian approximation in systems of
varying anharmonicity, we constructed a series of Morse potentials,
\begin{equation}
V(q)=V_{\text{eq}}+D_{e}[1-e^{-a(q-q_{\text{eq}})}]^{2}, \label{eq:morse}%
\end{equation}
with different values of the dissociation energy $D_{e}$ and anharmonicity
parameter $a$. In Eq.~(\ref{eq:morse}), $V_{\text{eq}}$ is the potential at
the equilibrium position $q_{\text{eq}}$. We chose to work in atomic units
($\hbar=1$) and mass-scaled coordinates. The initial wavepacket
was a real Gaussian with zero position and momentum, and with a width matrix
$A_{0}=\omega_{0}/2$ corresponding to the ground vibrational state of a
harmonic oscillator with frequency $\omega_{0}=0.00456\ \text{a.u.}%
=1000\ $cm$^{-1}$. The Morse parameters were $V_{\text{eq}}=0.1$ and
$q_{\text{eq}}=\sqrt{2/\omega_{0}}=20.95\ \text{a.u.}$. We also fixed the
global harmonic potential fitted to the Morse potentials at the equilibrium
position $q_{\text{eq}}$; its frequency,
\begin{equation}
\omega_{\text{eq}}=\sqrt{V^{\prime\prime}(q_{\text{eq}})}=\sqrt{2D_{e}a^{2}},
\label{eq:omega_0_morse}%
\end{equation}
was set to $0.0041\ \text{a.u.}=900\ $cm$^{-1}$. Anharmonicity of the
potential was controlled through the dimensionless constant
\begin{equation}
\chi=\frac{\omega_{\text{eq}}}{4D_{e}}. \label{eq:chi_morse}%
\end{equation}
Then, the $D_{e}$ and $a$ parameters were uniquely defined as
\begin{align}
D_{e}  &  =\frac{\omega_{\text{eq}}}{4\chi},\label{eq:d_morse}\\
a  &  =\sqrt{2\omega_{\text{eq}}\chi}. \label{eq:a_morse}%
\end{align}

The transition dipole moment was set to $1$. The wavepacket was always
propagated for $4000$ steps of $8\ $a.u.$\ \approx0.194\ $fs. Spectra
evaluated with the thawed-Gaussian, global harmonic, and single-Hessian
approaches discussed in Section~\ref{subsec:refhess} were compared with the
exact quantum dynamics calculations, obtained with the second-order
split-operator method. The position grid for the exact quantum dynamics
consisted of $16384$ points between $-200$ and $200$ atomic units. To avoid
artifacts of the finite-time calculation, all correlation functions were
multiplied by a Gaussian damping function corresponding to the Gaussian
broadening with half-width at half-maximum of $115\ $cm$^{-1}$. Spectra were
then computed from Eq.~(\ref{eq:sigma_abs}) and scaled according to the
maximum intensity.

\subsection{\label{subsec:otf_comp}On-the-fly \emph{ab initio} calculations}

The on-the-fly \emph{ab initio} implementation of the thawed Gaussian
approximation has been detailed in
Refs.~\onlinecite{Wehrle_Vanicek:2014, Wehrle_Vanicek:2015, Patoz_Vanicek:2018,Begusic_Vanicek:2018}.
Briefly, the method evaluates the required potential information along the
trajectory from an \emph{ab initio} electronic structure program. Our in-house
code performs the dynamics, transformation between Cartesian and normal-mode
coordinates, and interpolation of the Hessians if they are not computed at
each step (see Refs.~\onlinecite{Wehrle_Vanicek:2014,Patoz_Vanicek:2018}).

For ammonia, the \emph{ab initio} calculations were performed using the
complete active-space second-order perturbation theory, CASPT2(8/8), in
combination with the aug-cc-pVTZ basis set, as implemented in the Molpro2012.1
package.\cite{Werner_Schutz:2012, MOLPRO:2012} For the quinquethiophene, the
ground-state potential data were evaluated using the density functional
theory, while the time-dependent density functional theory was used for
geometry optimization and Hessian calculations in the first excited electronic
state; the functional was B3LYP and the basis set 6-31+G(d,p), as implemented
in Gaussian09.\cite{g09} All trajectories were propagated using a time step of
$8\ $a.u. for $1000$ steps in ammonia and for $997$ steps in quinquethiophene.
In ammonia the Hessian was computed at each step, whereas in quinquethiophene
the Hessian was evaluated only every four steps and interpolated in between;
such an interpolation was previously validated in
Ref.~\onlinecite{Wehrle_Vanicek:2014}. Before computing the spectra, the
correlation functions were multiplied with a Gaussian damping function
corresponding to the spectral Gaussian broadening with half-width at
half-maximum of $200\ $cm$^{-1}$. Further computational details about the
ammonia absorption spectrum can be found in
Ref.~\onlinecite{Wehrle_Vanicek:2015} and, about the quinquethiophene emission
spectrum, in Ref.~\onlinecite{Wehrle_Vanicek:2014}.

\section{\label{sec:resanddisc}Results and discussion}

\subsection{\label{subsec:morse_res}Morse potential}

\begin{figure}
[ptb]%
\includegraphics[width=1\textwidth]{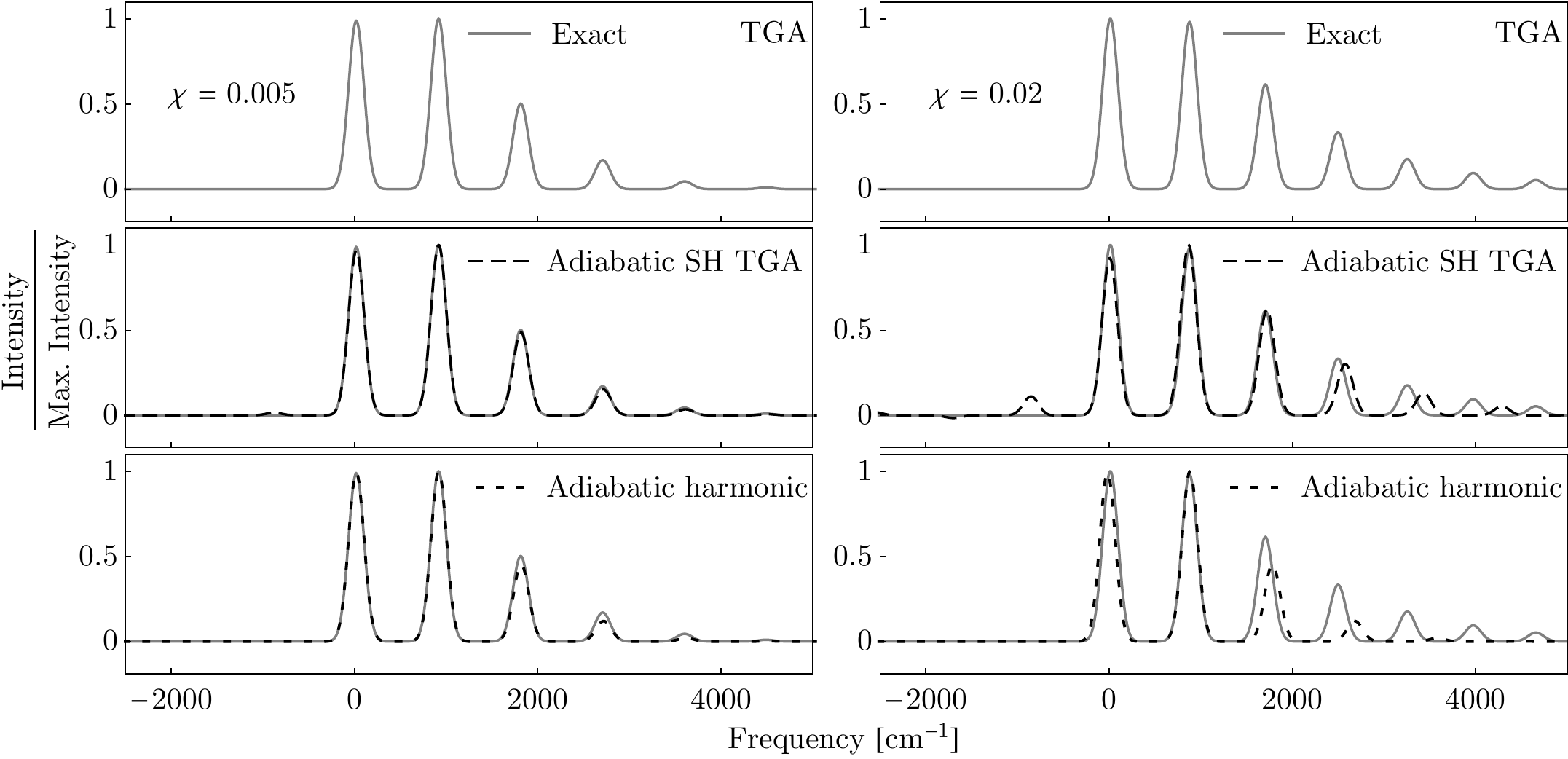}\caption{Spectra of two
Morse potentials with different anharmonicity constants $\chi$ evaluated using
the exact quantum dynamics, thawed Gaussian approximation (TGA, top),
adiabatic single-Hessian thawed Gaussian approximation (SH TGA, middle), and
adiabatic harmonic model (bottom). Left: $\chi=0.005$. Right: $\chi=0.02$. All
spectra were shifted to give the best overlap with the exact calculation and
the zero frequency was set to the 0--0 transition, i.e., the first peak of the
progression. All approximate spectra in the left panels overlap
almost perfectly with the exact spectrum, whereas larger differences between
the exact and approximate spectra are observed in the more anharmonic Morse
potential in the right-hand panels.}\label{fig:MorseSpectra}
\end{figure}

Figure~\ref{fig:MorseSpectra} compares the exact spectra of two Morse
potentials of different degrees of anharmonicity with those evaluated using
the standard thawed Gaussian approximation, its adiabatic single-Hessian
version, and the adiabatic harmonic method. In the weakly anharmonic potential
(Fig.~\ref{fig:MorseSpectra}, left), all methods perform well, with only the
global harmonic spectrum deviating slightly from the exact solution. In
contrast, in the more anharmonic Morse potential, the adiabatic harmonic model
recovers only the first few peaks. Interestingly, the single-Hessian version
seems to be more accurate than the standard thawed Gaussian approximation in
describing peak intensities.

\begin{figure}
\includegraphics[width=0.45\textwidth]{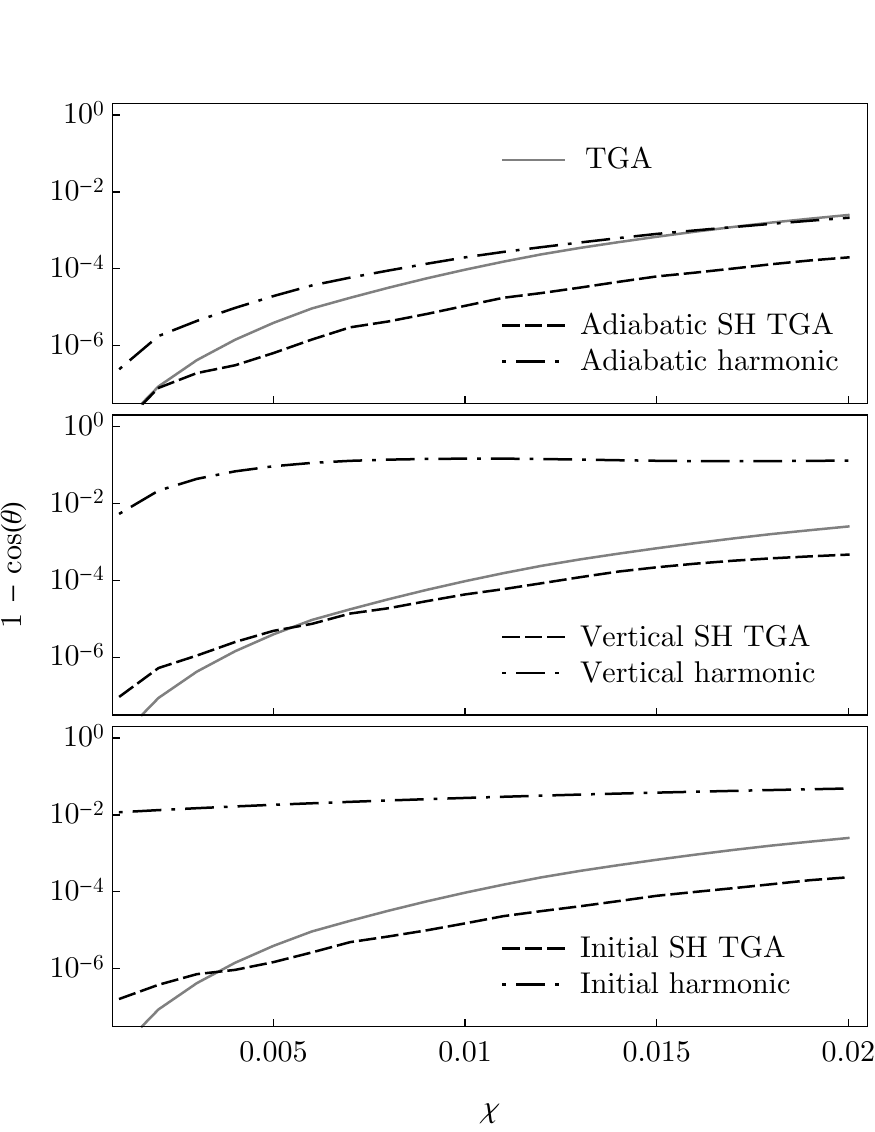} \caption{\label{fig:SpectralContrastAngle}Spectral contrast angles for Morse potentials with different anharmonicity constants $\chi$. The angles [Eq.~(\ref{eq:cos_theta})] compare approximate spectra evaluated using the thawed Gaussian approximation (TGA), its single-Hessian (SH) versions, and global harmonic methods with the exact spectrum. The single-Hessian and global harmonic results are presented for three different choices of the reference Hessian: adiabatic, vertical, and initial.}
\end{figure}

To quantify the accuracy of the approaches discussed in
Section~\ref{subsec:refhess}, we introduce the spectral contrast angle
$\theta$ between a reference ($\sigma_{\text{ref}}$) and approximate ($\sigma
$) spectra, conveniently defined through its cosine
\begin{equation}
\cos\theta= \frac{\sigma_{\text{ref}} \cdot\sigma}{\lVert\sigma_{\text{ref}%
}\rVert\lVert\sigma\rVert}, \label{eq:cos_theta}%
\end{equation}
with the inner product $\sigma_{1} \cdot\sigma_{2} = \int d\omega\sigma_{1}
(\omega) \sigma_{2} (\omega)$ of two spectra and norm of a spectrum
$\lVert\sigma\rVert= \sqrt{\sigma\cdot\sigma}$. Spectra evaluated with the
exact quantum dynamics are used as reference. In \emph{ab initio}
calculations, the errors in the absolute frequency shift of the spectrum
originate mostly from the limited accuracy of the electronic structure
methods. Therefore, even in the Morse potential, we first maximize the overlap
with the reference by shifting the computed spectra in frequency and then
evaluate the spectral contrast angle. The maximum overlap is found by scanning
through all possible shifts, with the increment determined by the numerical
resolution of the spectrum.

As shown in Fig.~\ref{fig:SpectralContrastAngle}, the accuracies of all
presented methods decrease with increasing anharmonicity of the potential.
However, the methods based on the thawed Gaussian approximation clearly
perform better than the global harmonic approaches. Moreover, the
single-Hessian results are nearly the same for all three choices of the
Hessian, which is not the case for the global harmonic approximations. The
errors in the spectra of more anharmonic potentials (see
Fig.~\ref{fig:MorseSpectra}) are reflected mainly in incorrect peak spacings,
which are almost exclusively determined by the classical trajectory guiding
the thawed Gaussian wavepacket---therefore, in the single-Hessian thawed
Gaussian approximation the choice of the Hessian affects the result only weakly.

\begin{figure}
\includegraphics[width=0.45\textwidth]{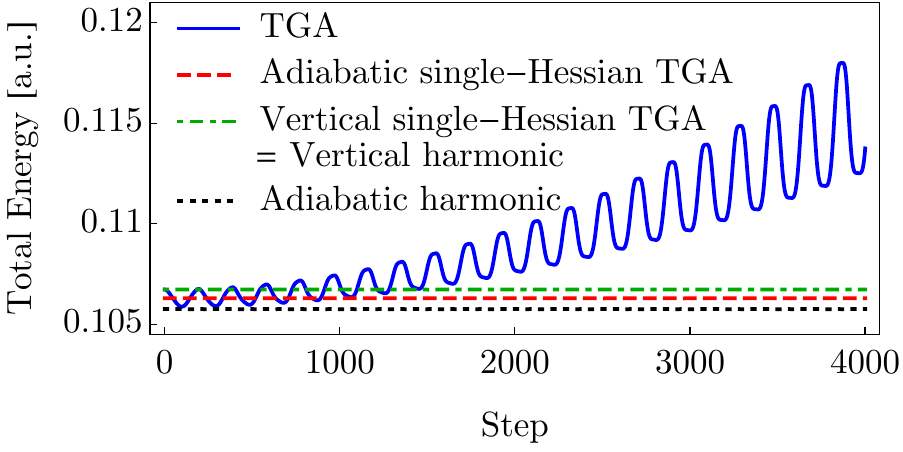}
\caption{\label{fig:Morse_EnergyConservation}Total energy of the wavepackets propagated in a Morse potential
($\chi = 0.005$, see Sec.~\ref{subsec:morse_comp}) using the thawed Gaussian approximation (TGA),
two single-Hessian approaches, and two harmonic models. For the initial single-Hessian thawed Gaussian
approximation (not shown for clarity), the energy is a horizontal line between those corresponding to the adiabatic
and vertical single-Hessian approaches.}
\end{figure}

Negative intensities in the spectra computed with the thawed Gaussian
approximation further increase the errors measured by the spectral contrast
angle. Such features are nearly eliminated in the single-Hessian version of
the thawed Gaussian approximation, which conserves energy exactly (see
Fig.~\ref{fig:Morse_EnergyConservation}); however, negative intensities still
arise even in the single-Hessian method due to the time dependence of the
effective single-Hessian potential (\ref{eq:SHLHA}).

\subsection{\label{subsec:ammonia_res}Absorption spectrum of ammonia}

Ammonia is a prototypical example of a floppy system, i.e., a system
exhibiting large-amplitude motion. Electronic excitation to the first excited
state is accompanied by a significant displacement of the umbrella inversion
mode, allowing the generated wavepacket to visit anharmonic regions of the
excited-state potential energy surface. Due to the small size of the system,
rich nuclear dynamics, and available experimental data, the absorption,
emission, and photoelectron spectra of ammonia have served as benchmarks for
different methods built specifically to treat the anharmonicity
effects.\cite{Tang_Tannor:1990,Tang_Imre:1991,Capobianco_Peluso:2012,Baiardi_Barone:2017}
In particular, the on-the-fly \emph{ab initio} thawed Gaussian approximation
showed significant improvement over the global harmonic
models.\cite{Wehrle_Vanicek:2015}

\begin{figure}
[ptb]%
\includegraphics[width=1\textwidth]{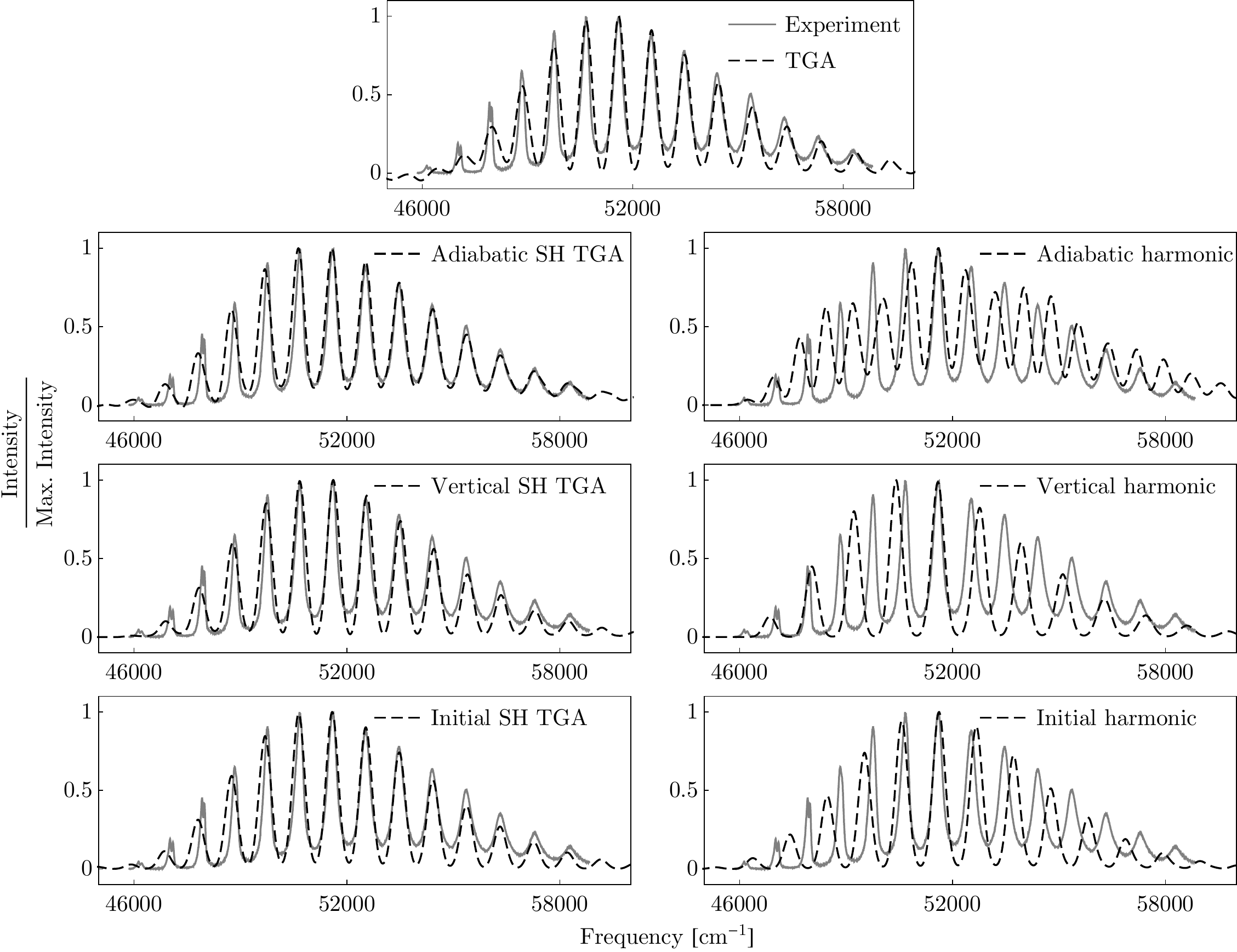}\caption{Experimental
absorption spectra of ammonia measured in gas phase at $175\ $K\cite{Chen_Caldwell:1999} compared with those evaluated using the on-the-fly
\emph{ab initio} thawed Gaussian approximation (TGA), its single-Hessian (SH)
version, and global harmonic models. The single-Hessian and global harmonic
results are presented for three different choices of the reference Hessian:
adiabatic, vertical, and initial. Computed spectra were scaled and shifted
according to the highest peak of the experiment.}\label{fig:Ammonia_Comparison}%

\end{figure}

Figure~\ref{fig:Ammonia_Comparison} compares the global harmonic and
single-Hessian approaches with the on-the-fly \emph{ab initio} thawed Gaussian
approximation\cite{Wehrle_Vanicek:2015} and with the experimental absorption
spectrum of ammonia.\cite{Chen_Caldwell:1999} All single-Hessian methods
recover both the peak positions and intensities of the standard thawed
Gaussian approximation. In contrast, all global harmonic models yield
different and rather inaccurate spectra. Most interesting are the adiabatic
single-Hessian thawed Gaussian approximation and adiabatic global harmonic
model: although both methods use only one (adiabatic) Hessian, the former
performs better than any other presented method, including the standard thawed
Gaussian approximation, whereas the latter performs the worst. These results
indicate that the single-Hessian thawed Gaussian approximation cannot be
discarded in advance based on the performance of global harmonic models; in
fact, its accuracy is much closer to that of the thawed Gaussian
approximation. Indeed, even the initial (ground-state) single Hessian approach
reproduces almost perfectly the result of the standard on-the-fly \emph{ab
initio} thawed Gaussian approximation.

\subsection{\label{subsec:quinquethiophene_res}Emission spectrum of
quinquethiophene}

Due to their potential in molecular electronics, polythiophenes and their
derivatives have been studied extensively. Oligothiophenes have also served as
a model system for studying the dependence of optical properties on the system
size. They present a challenge for computing vibrationally resolved electronic
spectra due to the torsional degrees of freedom, which cannot be treated with
global harmonic models. Wehrle et al.\cite{Wehrle_Vanicek:2014} showed that
the on-the-fly \emph{ab initio} thawed Gaussian approximation performs well
despite the double-well character of the potential along the torsional modes
connecting the planar and twisted structures.

\begin{figure}
[ptb]%
\includegraphics[width=1\textwidth]{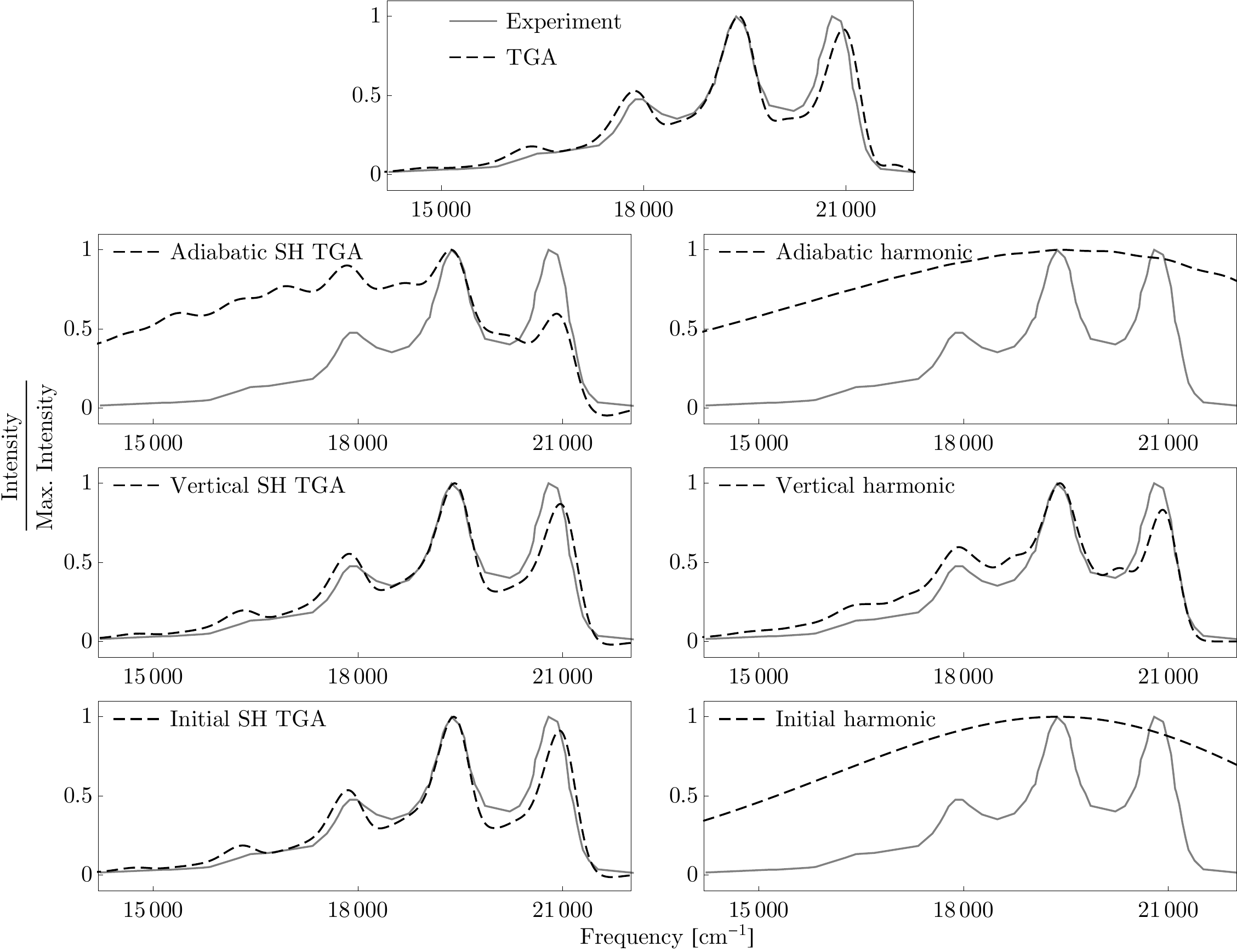}\caption{Analogous
to Fig.~\ref{fig:Ammonia_Comparison}, but for the emission spectrum of
quinquethiophene. The experiment was measured in ethanol glass at
$77\ $K.\cite{Becker_Elisei:1996}}\label{fig:Thio5_Comparison}
\end{figure}

In Fig.~\ref{fig:Thio5_Comparison}, we compare the
experimental\cite{Becker_Elisei:1996} emission spectrum of quinquethiophene,
an oligomer composed of five thiophene units, and corresponding spectra
computed with various approximations discussed in Section~\ref{sec:theory}.
The single-Hessian approaches using the initial (excited-state) and vertical
Hessians produce almost the same spectra as the standard thawed Gaussian
approximation\cite{Wehrle_Vanicek:2014} (shown in
Fig.~\ref{fig:Thio5_Comparison}, top). However, this is not the case for the
adiabatic single-Hessian method, which yields a broad spectrum due to the
incorrect description of the torsional degrees of freedom. As discussed in
Ref.~\onlinecite{Wehrle_Vanicek:2014}, the initial wavepacket is placed at the
top of a potential barrier along the torsional modes, which results in a
constant but slow wavepacket spreading. The adiabatic Hessian has all
frequencies positive and is therefore qualitatively inappropriate.
Interestingly, the initial single-Hessian approach, which propagates a frozen
Gaussian, results in a rather accurate spectrum, implying that the errors of
using the adiabatic Hessian arise due to the incorrect width of the Gaussian wavepacket.

In contrast, the failure of the adiabatic global harmonic model
(Fig.~\ref{fig:Thio5_Comparison}, top right) is not related to the Hessian,
but rather to the large displacement of the ground-state potential minimum
from the initial geometry. The computed emission spectrum is nearly
featureless because the wavepacket quickly drifts away from the initial planar
geometry and does not return during the short dynamics considered for spectra
simulations. This explanation is supported by the equally featureless spectrum
of the adiabatic shift model, i.e., the initial harmonic model (see
Fig.~\ref{fig:Thio5_Comparison}, bottom right), which has the same
displacement of the adiabatic global harmonic model but uses the initial
(excited-state) Hessian.

\section{Conclusion}

In conclusion, we have presented and validated an efficient method for
evaluating low-resolution vibronic spectra of polyatomic
molecules. The proposed single-Hessian thawed Gaussian approximation, whose
computational cost lies between those of the global harmonic and thawed
Gaussian approximations, performs surprisingly well, in some cases even better
than the more computationally demanding thawed Gaussian approximation.
Moreover, unlike the standard thawed Gaussian approximation, the
single-Hessian approach conserves total energy exactly. We have shown that
despite the conservation of energy, the computed spectra may still contain
negative intensities due to the time dependence of the effective Hamiltonian.
Yet, the negative spectral features are significantly smaller compared with
the standard thawed Gaussian approximation. In contrast to the spectra
evaluated using the global harmonic approaches, those computed with the
single-Hessian thawed Gaussian approximation depend only weakly on the
reference Hessian. Therefore, the single-Hessian approach offers a
considerable and systematic improvement over the commonly used global harmonic
models at the cost of a single \emph{ab initio} classical trajectory.

\begin{acknowledgments}
The authors acknowledge the financial support from the European Research
Council (ERC) under the European Union's Horizon 2020 research and innovation
programme (grant agreement No. 683069 -- MOLEQULE).
\end{acknowledgments}

\appendix

\section{\label{sec:app_HK}Single-Hessian approximations of the Herman--Kluk
prefactor}

Within the Herman--Kluk\cite{Herman_Kluk:1984,Herman:1986,Kay:1994a,
Kay:1994b} semiclassical initial value
representation,\cite{Miller:1970,Miller:2001,Kay:2005} the quantum evolution
operator is approximated as
\begin{equation}
e^{-i\hat{H}t/\hbar}\approx h^{-D}\int dq_{0}dp_{0}R_{t}(q_{0},p_{0}%
)e^{iS_{t}/\hbar}|q_{t}p_{t}\rangle\langle q_{0}p_{0}|, \label{eq:HK}%
\end{equation}
where $D$ is the number of degrees of freedom, $S_{t}:=\int_{0}^{t}%
L_{t^{\prime}}dt^{\prime}$ the classical action,
\begin{equation}
R_{t}(q_{0},p_{0})=\sqrt{\det\bigg[\frac{1}{2}(M_{qq}+\Gamma^{-1}\cdot
M_{pp}\cdot\Gamma-M_{qp}\cdot\Gamma-\Gamma^{-1}\cdot M_{pq})\bigg]}
\label{eq:HK_prefactor}%
\end{equation}
the Herman--Kluk prefactor, $M_{ab}=\partial a_{t}/\partial b_{0}$ components
of the stability matrix, $|q_{t}p_{t}\rangle$ the coherent state whose
wavefunction in position representation is
\begin{equation}
\langle q|q_{t}p_{t}\rangle=\left[  \det(\operatorname{Im}\Gamma/\pi
\hbar)\right]  ^{1/4}\exp\bigg[\frac{i}{\hbar}\left(  \frac{1}{2}(q-q_{t}%
)^{T}\cdot\Gamma\cdot(q-q_{t})+p_{t}^{T}\cdot(q-q_{t})\right)  \bigg],
\end{equation}
$\Gamma$ denotes a pure imaginary symmetric coherent state width matrix (i.e.,
$\Gamma^{\ast}=-\Gamma$ and $\Gamma^{T}=\Gamma$) , and $q_{t}$ and $p_{t}$
evolve classically according to Eqs.~(\ref{eq:q_dot})--(\ref{eq:p_dot}).

Reversing the main idea of the $PQ$ method\cite{Heller:1976} mentioned in
Section~\ref{subsec:H_map},\ in the log-derivative
formulation,\cite{Gelabert_Miller:2000,DiLiberto_Ceotto:2016} the Herman--Kluk
prefactor is expressed in terms of an auxiliary matrix
\begin{equation}
\alpha_{t}=P_{t}\cdot Q_{t}^{-1}=m\cdot\dot{Q}_{t}\cdot Q_{t}^{-1},
\label{eq:alpha_t}%
\end{equation}
where $Q_{t}=M_{qq}\cdot Q_{0}+M_{qp}\cdot P_{0}=M_{qq}-M_{qp}\cdot\Gamma$, as
\label{eq:Q_t}
\begin{equation}
R_{t}=\sqrt{\det\bigg[\frac{1}{2}(I_{D}+\alpha_{0}^{-1}\cdot\alpha_{t}%
)\bigg]}\exp\bigg[\frac{1}{2}\int_{0}^{t}dt^{\prime}\text{Tr}(m^{-1}%
\cdot\alpha_{t^{\prime}})\bigg]. \label{eq:HK_prefactor_log}%
\end{equation}
Matrix $Q_{t}$ defined here is equivalent to that of Eq.~(\ref{eq:A_QP}) for a
specific choice of initial conditions: $Q_{0}=I_{D}$ and $P_{0}=-\Gamma$ (see
Appendix~\ref{sec:app_H_map}). Matrix $\alpha_{t}$, whose initial value is
$\alpha_{0}=-\Gamma$, obeys the same equation of motion as the matrix $A_{t}$
of the thawed Gaussian approximation [Eq.~(\ref{eq:A_dot})]; the connection
between $\alpha_{t}$ and $A_{t}$ was discussed, e.g., in Ref.~\onlinecite{Gelabert_Miller:2000}.

To compare different approximations to the prefactor $R_{t}$ with the
single-trajectory single-Hessian thawed Gaussian approximation, we consider
only a single trajectory in Eq.~(\ref{eq:HK}) and approximate the propagated
wavepacket as
\begin{equation}
e^{-i\hat{H}t/\hbar}|\psi_{i}\rangle\approx R_{t}e^{iS_{t}/\hbar}|q_{t}%
p_{t}\rangle. \label{eq:HK_single}%
\end{equation}
Then, $\Gamma=A_{0}$ of the initial wavepacket $\psi_{i}$ and the wavepacket
at time $t$ is a Gaussian (\ref{eq:GWP}) with parameters $A_{t}=A_{0}$ and
$\gamma_{t}$ given by
\begin{equation}
e^{i\gamma_{t}/\hbar}=R_{t}e^{iS_{t}/\hbar}.
\end{equation}

In what follows, we apply the single-Hessian potential [Eq.~(\ref{eq:SHLHA})]
to the Herman--Kluk prefactor and its approximations. For a constant Hessian,
assuming for simplicity that $\Gamma$, $m$, and $V_{\text{ref}}^{\prime\prime
}(q_{\text{ref}})$ commute (which is valid, e.g., if $D=1$, or if all three
matrices are diagonal, or if spherical Gaussians and mass-scaled coordinates
are used, i.e., $\Gamma\propto I_{D}$ and $m\propto I_{D}$), the Herman--Kluk
prefactor simplifies to\cite{Gelabert_Miller:2000}
\begin{equation}
R_{t}=\exp\bigg[\frac{1}{2}\int_{0}^{t}dt^{\prime}\text{Tr}(m^{-1}\cdot
\tilde{\alpha}_{t^{\prime}})\bigg]\label{eq:HK_prefactor_SH}%
\end{equation}
Matrix $\tilde{\alpha}_{t}$ evolves as $A_{t}$ of the single-Hessian thawed
Gaussian approximation, but with a modified initial condition
\begin{equation}
\tilde{\alpha}_{0}=-\frac{1}{2}(A_{0}+m\cdot A_{0}^{-1}\cdot A_{\text{ref}%
}\cdot m^{-1}\cdot A_{\text{ref}}),\label{eq:alpha_tilde_0}%
\end{equation}
where $A_{\text{ref}}$ corresponds to the coherent state of a harmonic potential with force
constant matrix $k=V_{\text{ref}}^{\prime\prime}(q_{\text{ref}})$, i.e.,
\begin{equation}
A_{\text{ref}}\cdot m^{-1}\cdot A_{\text{ref}}=-V_{\text{ref}}^{\prime\prime
}(q_{\text{ref}}).\label{eq:A_ref}%
\end{equation}
Equation~(\ref{eq:HK_prefactor_SH}) coincides with the slowly varying Hessian
approximation of Gelabert et al.\cite{Gelabert_Miller:2000}; however, their
approximation formally assumes a time-dependent Hessian for the evolution of
$\tilde{\alpha}_{t}$, whereas here, Eq.~(\ref{eq:HK_prefactor_SH}) is an exact
expression for the prefactor in the approximate potential (\ref{eq:SHLHA}).
Because matrix $\tilde{\alpha}_{t}$ is, in general, complex at $t>0$, the norm
of the \textquotedblleft single-Hessian Herman--Kluk\textquotedblright%
\ wavepacket is not conserved. This is remedied easily by taking only the
imaginary part of $\tilde{\alpha}_{t}$ in Eq.~(\ref{eq:HK_prefactor_SH}), or,
equivalently, by renormalizing the wavepacket at each step.

In Johnson's multichannel Wentzel--Kramers--Brillouin
approximation,\cite{Gelabert_Miller:2000,Issack_Roy:2005,Issack_Roy:2007,Issack_Roy:2007a}
one assumes that the matrix $\alpha_{t}$ varies slowly, i.e., $\dot{\alpha
}_{t}\approx0$, which yields
\begin{align}
R_{t}  &  =\exp\bigg\lbrace-\frac{i}{2}\int_{0}^{t}dt^{\prime}%
\operatorname{Tr}\big[\big(m^{-1}\cdot V^{\prime\prime}(q_{t^{\prime}%
})\big)^{1/2}\big]\bigg\rbrace\label{eq:HK_prefactor_J_intermediate}\\
&  =\exp\bigg[-\frac{i}{\hbar}\int_{0}^{t}dt^{\prime}\sum_{j=1}^{D}\frac{1}%
{2}\hbar\omega_{j}(t^{\prime})\bigg], \label{eq:HK_prefactor_J}%
\end{align}
where $\omega_{j}(t)$ are time-dependent frequencies obtained from the Hessian
evaluated at $q_{t}$. This method involves a time-dependent Hessian and is,
therefore, closer to the original thawed Gaussian approximation than to its
single-Hessian version. However, if Johnson's approximation is combined with
the single-Hessian potential~(\ref{eq:SHLHA}), the time-dependent frequencies
$\omega_{j}(t)$ are replaced by the reference frequencies $\omega
_{\text{ref},j}$ obtained from the reference Hessian $V_{\text{ref}}%
^{\prime\prime}(q_{\text{ref}})$ and the integral in
Eq.~(\ref{eq:HK_prefactor_J}) is trivial.

The adiabatic approximation\cite{Guallar_Miller:1999,Guallar_Miller:2000} of
the Herman--Kluk prefactor assumes an instantaneously diagonal Hessian at each
time step, i.e., it neglects the offdiagonal entries of the full Hessian
matrix. Within the single-Hessian approximation, the resulting expression for
$R_{t}$ is the same as for the single-Hessian Herman--Kluk
[Eq.~(\ref{eq:HK_prefactor_SH})] except for a modified (diagonal) Hessian.

Finally, the crudest approximation is to replace the prefactor by unity, which
is known as the prefactor-free approach;\cite{Tatchen_Pollak:2009} then,
$\gamma_{t}=S_{t}$ and no Hessian computation is needed.

\begin{table}
[ptb]%
\caption{Equations of motion for parameters $A_{t}$ and $\gamma_{t}$ of the
Gaussian wavepacket (\ref{eq:GWP}) propagated with the single-Hessian thawed
Gaussian approximation, exact for the approximate Hamiltonian (\ref{eq:SHLHA}), and with the single-Hessian and single-trajectory versions of the
approximations discussed in Appendix~\ref{sec:app_HK}. $\tilde{\alpha}_{0}$ is defined in
Eq.~(\ref{eq:alpha_tilde_0}), $A_{\text{ref}}$ in Eq.~(\ref{eq:A_ref}), and we use $-(i\hbar/2) \text{Tr}(m^{-1}\cdot
A_{\text{ref}})=\sum_{j=1}^{D}\frac{1}{2}\hbar\omega_{\text{ref},j}$. For the single-Hessian Herman--Kluk approximation, we assume that matrices $\Gamma$, $m$, and $V^{\prime \prime}_{\text{ref}}(q_{\text{ref}})$ commute. For the
frozen Gaussian approximation,\cite{Heller:1981,Davis_Heller:1981} the general
expression $\dot{\gamma}_{t}=p_{t}^{T}\cdot m^{-1}\cdot p_{t}-\langle\hat
{H}\rangle$ is expanded using the single-Hessian potential~(\ref{eq:SHLHA})
and the total energy of a Gaussian wavepacket [Eqs.~(\ref{eq:E_tot})--(\ref{eq:E_scl})] applied to a coherent state ($A_{t} = A_{0}$).}\label{tab:TGA_HK_eq}%
\begin{ruledtabular}
\begin{tabular}[t]{l l}
Thawed Gaussian & Frozen Gaussian \\
\hline
$\begin{aligned}
\dot{A}_{t}  &  =- A_{t}\cdot m^{-1}\cdot A_{t} - V^{\prime\prime}_{\text{ref}}(q_{\text{ref}})\\
\dot{\gamma}_{t}  &  =L_{t}+\frac{i\hbar}{2}\text{Tr}\left(  m^{-1}\cdot A_{t}\right)
\end{aligned}$
&
$\begin{aligned}
\dot{A}_{t} & = 0 \\
\dot{\gamma}_{t}  &  =L_{t} - \frac{i\hbar}{2} \text{Tr}( m^{-1} \cdot \tilde{\alpha}_{0})
\end{aligned}$
\\
\hline
Herman--Kluk & Adiabatic Herman--Kluk \\
\hline
$\begin{aligned}
&\dot{A}_{t}    = 0  \\
&\dot{\gamma}_{t}    = L_{t}-\frac{i\hbar}{2}\text{Tr}\left(  m^{-1}\cdot \tilde{\alpha}_{t} \right) \\
&\dot{\tilde{\alpha}}_{t}   = - \tilde{\alpha}_{t}\cdot m^{-1}\cdot \tilde{\alpha}_{t} - V^{\prime\prime}_{\text{ref}}(q_{\text{ref}})
\end{aligned}$
&
$\begin{aligned}
&\dot{A}_{t}   = 0  \\
&\dot{\gamma}_{t}   = L_{t}-\frac{i\hbar}{2}\text{Tr}\left(  m^{-1}\cdot \tilde{\alpha}_{t} \right) \\
&\dot{\tilde{\alpha}}_{t,jj}   =-\frac{\tilde{\alpha}_{t,jj}^2}{m_j} - m_j \omega_{\text{ref},j}^2
\end{aligned}$
\\
\hline
Johnson  & Prefactor-free \\
\hline
$\begin{aligned}
\dot{A}_{t}  &  = 0 \\
\dot{\gamma}_{t}  & = L_{t} + \frac{i\hbar}{2} \text{Tr} \big( m^{-1} \cdot A_{\text{ref}} \big)
\end{aligned}$
&
$\begin{aligned}
\dot{A}_{t}  &  = 0 \\
\dot{\gamma}_{t}  & = L_{t}
\end{aligned}$
\end{tabular}
\end{ruledtabular}
\end{table}

Equations of motion for parameters $A_{t}$ and $\gamma_{t}$ in the
single-Hessian thawed Gaussian, Herman--Kluk, Johnson's, adiabatic
Herman--Kluk, and prefactor-free approximations are summarized in
Table~\ref{tab:TGA_HK_eq}, where we also present analogous expressions for
Heller's frozen Gaussian approximation.\cite{Heller:1981} Single-Hessian
thawed Gaussian wavepacket has a time-dependent width, whereas the other
approximations propagate a coherent state with only a modified phase factor.
Special case is the initial single-Hessian approach, which uses the initial
Hessian for the single-Hessian thawed Gaussian propagation. Then, $\dot{A}%
_{t}=0$ holds even for the thawed Gaussian wavepacket and $\gamma_{t}$ is the
same for the thawed Gaussian, Herman--Kluk, Johnson's, and frozen Gaussian
approximations. Let us emphasize that in the multiple-trajectory
implementations of the single-Hessian Herman--Kluk, Johnson's, and frozen
Gaussian methods, $\Gamma$ is a free parameter; for $\Gamma=A_{\text{ref}}$,
the three approximations are equivalent. In contrast, in the single-trajectory
thawed Gaussian approximation, because the initial width parameter $A_{0}$ is
fixed by the wavepacket $\psi_{i}$, using a constant Hessian does not imply a
time-independent matrix $A_{t}$. Therefore, the single-Hessian method is,
despite similarities, fundamentally different from other approaches.

\begin{table}
\caption{\label{tab:Morse_TGA_HK}Cosines of the spectral contrast angles [Eq.~(\ref{eq:cos_theta})]
comparing the exact spectrum of a Morse potential ($\chi = 0.02$, see Sec.~\ref{subsec:morse_comp}) with the
spectra evaluated using the single-Hessian thawed Gaussian approximation (SH TGA) and single-Hessian
single-trajectory approximations discussed in Appendix~\ref{sec:app_HK}. In a one-dimensional system, the adiabatic
Herman--Kluk approximation is equivalent to the Herman--Kluk method. Results for adiabatic, vertical, and
initial reference Hessians are shown. The top three rows contain contrast angles of the spectra shifted so that their
overlaps with the exact (reference) spectrum are maximal; the rows below refer to the unshifted spectra, where the
errors due to constant horizontal shifts of the spectra are accounted for.}
\begin{ruledtabular}
\begin{tabular}{lccccc}
Reference &SH&Herman&Johnson&Frozen&Prefactor\\
Hessian &TGA&Kluk&       &Gaussian&free\\
\hline
&&&Shifted&&\\
\hline
Adiabatic&0.975&0.973&0.973&0.973&0.973\\
Vertical&0.964&0.973&0.973&0.973&0.973\\
Initial&0.973&0.973&0.973&0.973&0.973\\
\hline
&&&Not shifted&&\\
\hline
Adiabatic&0.975&0.973&0.973&0.973&0.006\\
Vertical&0.242&0.243&0.243&0.172&0.006\\
Initial&0.899&0.899&0.899&0.899&0.006
\end{tabular}
\end{ruledtabular}

\end{table}

Various single-Hessian approaches are compared numerically in
Table~\ref{tab:Morse_TGA_HK}. The results confirm that the single-Hessian
thawed Gaussian approximation is not identical to the single-trajectory
Herman-Kluk propagator or any of its several simplified versions. That the
differences between the methods are only small may be attributed to a weak
distortion of the model system---greater difference between the ground- and
excited-state Hessians would lead to greater deformations of the wavepacket,
which cannot be described by a single coherent state [Eq.~(\ref{eq:HK_single}%
)]. The shifted spectra obtained with single-Hessian Johnson's, frozen
Gaussian, and prefactor-free approximations are the same because the methods
differ only by a factor $\exp(it\Delta)$, where $\Delta$ is a real constant
depending on the methods that are compared (see Table~\ref{tab:TGA_HK_eq}).
Finally, all methods except for the prefactor-free approximation yield exactly
the same result if the initial-state Hessian is used as a reference, in
agreement with the theoretical justification given above.

The single-Hessian approximations of the coherent-state methods are not
necessarily useful in practice and are presented here only for comparison with
the single-Hessian thawed Gaussian approximation. Indeed, in the usual
multi-trajectory setup, the single-Hessian Herman--Kluk approach, which is
equivalent to the harmonic approximation mentioned briefly in
Ref.~\onlinecite{DiLiberto_Ceotto:2016}, would already be a feasible
computational method and no further approximations of the prefactor would be
needed. Otherwise, approaches based on the
Herman--Kluk\cite{Ianconescu_Pollak:2013,Bonfanti_Pollak:2018} and
Johnson's\cite{Issack_Roy:2005,Issack_Roy:2007,Issack_Roy:2007a}
approximations have been validated on difficult systems, where accurate
calculations require the evaluation of Hessians along each trajectory.

\section{\label{sec:app_H_map}Energy of the Gaussian wavepacket and the
mapping Hamiltonian}

\subsection{\label{subsec:app_relations}Useful relations}

Auxiliary matrices $Q_{t}$ and $P_{t}$, defined by Eqs.~(\ref{eq:A_QP}) and
(\ref{eq:P_t}), satisfy the relations\cite{Lee_Heller:1982,Faou_Lubich:2009}
\begin{align}
Q_{t}^{T}\cdot P_{t}-P_{t}^{T}\cdot Q_{t}  &  =0,\label{eq:QP_transpose}\\
Q_{t}^{\dagger}\cdot P_{t}-P_{t}^{\dagger}\cdot Q_{t}  &  =2iQ_{0}^{\dagger
}\cdot\operatorname{Im}A_{0}\cdot Q_{0}. \label{eq:QP_Hermitian}%
\end{align}
The former is obtained from $P_{t}^{T}\cdot Q_{t}=(Q_{t}^{T}\cdot P_{t})^{T}$
using Eq.~(\ref{eq:A_QP}) for $P_{t}$, the latter by showing that the time
derivative of the left hand side is zero and by confirming the relation at
time zero---by realizing that
\[
P_{0}^{\dagger}\cdot Q_{0}=(Q_{0}^{\dagger}\cdot P_{0})^{\dag}=(Q_{0}^{\dag
}\cdot A_{0}\cdot Q_{0})^{\dag}=Q_{0}^{\dag}\cdot A_{0}^{\ast}\cdot Q_{0}.
\]
A remarkable relation\cite{Hagedorn:1998,Faou_Lubich:2009}
\begin{equation}
\operatorname{Im}(P_{t}\cdot Q_{t}^{-1})=(Q_{t}^{\dagger})^{-1}\cdot
Q_{0}^{\dag}\cdot\operatorname{Im}A_{0}\cdot Q_{0}\cdot Q_{t}^{-1}
\label{eq:Im_PQinverse}%
\end{equation}
can be deduced from Eqs.~(\ref{eq:QP_transpose}) and (\ref{eq:QP_Hermitian}):
\begin{align}
2i\operatorname{Im}(P_{t}\cdot Q_{t}^{-1})  &  =P_{t}\cdot Q_{t}^{-1}%
-P_{t}^{\ast}\cdot(Q_{t}^{\ast})^{-1}\label{eq:Im_PQinverse_der1}\\
&  =(Q_{t}^{\dagger})^{-1}\cdot\lbrack Q_{t}^{\dagger}\cdot P_{t}%
-Q_{t}^{\dagger}\cdot P_{t}^{\ast}\cdot(Q_{t}^{\ast})^{-1}\cdot Q_{t}]\cdot
Q_{t}^{-1}\label{eq:Im_PQinverse_der2}\\
&  =(Q_{t}^{\dagger})^{-1}\cdot\lbrack Q_{t}^{\dagger}\cdot P_{t}%
-P_{t}^{\dagger}\cdot Q_{t}^{\ast}\cdot(Q_{t}^{\ast})^{-1}\cdot Q_{t}]\cdot
Q_{t}^{-1}\label{eq:Im_PQinverse_der3}\\
&  =2i(Q_{t}^{\dagger})^{-1}\cdot Q_{0}^{\dag}\cdot\operatorname{Im}A_{0}\cdot
Q_{0}\cdot Q_{t}^{-1}. \label{eq:Im_PQinverse_der4}%
\end{align}
Equation (\ref{eq:Im_PQinverse_der3}) follows from
Eq.~(\ref{eq:Im_PQinverse_der2}) because $Q_{t}^{\dagger}\cdot P_{t}^{\ast
}=(Q_{t}^{T}\cdot P_{t})^{\ast}=(P_{t}^{T}\cdot Q_{t})^{\ast}=P_{t}^{\dagger
}\cdot Q_{t}^{\ast}$, where we used Eq.~(\ref{eq:QP_transpose}), and
Eq.~(\ref{eq:Im_PQinverse_der4}) follows from (\ref{eq:Im_PQinverse_der3}) by
applying Eq.~(\ref{eq:QP_Hermitian}).

\subsection{\label{subsec:app_energy}Energy of the thawed Gaussian wavepacket}

The total energy of the thawed Gaussian wavepacket, computed as the
expectation value $E=\langle\psi(t)|\hat{H}_{\text{eff}}(t)|\psi(t)\rangle$,
can be split as\cite{Wehrle_Vanicek:2015}
\begin{equation}
E=E_{\text{cl}}+E_{\text{sc}} \label{eq:E_tot}%
\end{equation}
into the \textquotedblleft classical\textquotedblright\ energy of the central
trajectory,
\begin{equation}
E_{\text{cl}}=\frac{1}{2}p_{t}^{T}\cdot m^{-1}\cdot p_{t}+V(q_{t})=H\left(
q_{t},p_{t}\right)  , \label{eq:E_cl}%
\end{equation}
and \textquotedblleft semiclassical\textquotedblright\ energy
\begin{equation}
E_{\text{sc}}=\frac{1}{4}\hbar\text{Tr}\bigg[\left(  A_{t}\cdot m^{-1}\cdot
A_{t}^{\ast}+V^{\prime\prime}(q_{t})\right)  \cdot\left(  \text{Im}%
A_{t}\right)  ^{-1}\bigg]. \label{eq:E_scl}%
\end{equation}

The first factor inside the trace can be rewritten as
\begin{align}
&  A_{t}\cdot m^{-1}\cdot A_{t}^{\ast}+V^{\prime\prime}(q_{t})\nonumber\\
&  =A_{t}^{T}\cdot m^{-1}\cdot A_{t}^{\ast}+V^{\prime\prime}(q_{t}%
)\label{eq:H_der1}\\
&  =(Q_{t}^{T})^{-1}\cdot(P_{t}^{T}\cdot m^{-1}\cdot P_{t}^{\ast}+Q_{t}%
^{T}\cdot V^{\prime\prime}(q_{t})\cdot Q_{t}^{\ast})\cdot(Q_{t}^{\ast}%
)^{-1}\label{eq:H_der2}\\
&  =2(Q_{t}^{T})^{-1}\cdot\mathcal{H}_{\text{sc}}(Q_{t},P_{t},q_{t})^{\ast
}\cdot(Q_{t}^{\ast})^{-1}\label{eq:H_der3}\\
&  =2(Q_{t}^{\dag})^{-1}\cdot\mathcal{H}_{\text{sc}}(Q_{t},P_{t},q_{t})\cdot
Q_{t}^{-1}. \label{eq:H_der4}%
\end{align}
Equation~(\ref{eq:H_der1}) holds because $A_{t}$ is symmetric, in
Eq.~(\ref{eq:H_der2}) we used expression~(\ref{eq:A_QP}) for $A_{t}$, and in
Eq.~(\ref{eq:H_der3}) we introduced a matrix-valued function
\begin{equation}
\mathcal{H}_{\text{sc}}(Q_{t},P_{t},q_{t})=\frac{1}{2}P_{t}^{\dagger}\cdot
m^{-1}\cdot P_{t}+\frac{1}{2}Q_{t}^{\dagger}\cdot V^{\prime\prime}(q_{t})\cdot
Q_{t}. \label{eq:calH_sc}%
\end{equation}
The last step (\ref{eq:H_der4}) follows because both $E_{\text{sc}}$ and
$\operatorname{Im}A_{t}$ in Eq.~(\ref{eq:E_scl}) are real. As for the second
factor inside the trace in Eq.~(\ref{eq:E_scl}), relations~(\ref{eq:A_QP}) and
(\ref{eq:Im_PQinverse}) imply that
\begin{equation}
\left(  \text{Im}A_{t}\right)  ^{-1}=Q_{t}\cdot(Q_{0}^{\dag}\cdot
\operatorname{Im}A_{0}\cdot Q_{0})^{-1}\cdot Q_{t}^{\dagger}.
\label{eq:Re_A_v2}%
\end{equation}
Substitution of expressions~(\ref{eq:Re_A_v2}) and (\ref{eq:H_der3}) for the
two factors into the relation~(\ref{eq:E_scl}) for the semiclassical energy
gives
\begin{equation}
E_{\text{sc}}=\frac{1}{2}\hbar\text{Tr}[\mathcal{H}_{\text{sc}}(Q_{t}%
,P_{t},q_{t})\cdot(Q_{0}^{\dagger}\cdot\operatorname{Im}A_{0}\cdot Q_{0}%
)^{-1}]. \label{eq:E_der3}%
\end{equation}

The choice of $Q_{0}$ is not determined by the definitions (\ref{eq:A_QP}) and
(\ref{eq:P_t})\ of $Q_{t}$ and $P_{t}$. A common choice is $Q_{0}=I_{D}$ (a
$D$-dimensional identity matrix) and $P_{0}=A_{0}$, which yields
\begin{equation}
E_{\text{sc}}=\frac{1}{2}\hbar\text{Tr}[\mathcal{H}_{\text{sc}}(Q_{t}%
,P_{t},q_{t})\cdot(\operatorname{Im}A_{0})^{-1}]. \label{E_scl_final_v1}%
\end{equation}

However, one can remove all constant factors from Eq.~(\ref{eq:E_der3}) by
setting $Q_{0}=(2\operatorname{Im}A_{0}/\hbar)^{-1/2}\cdot U$, with an
arbitrary unitary matrix $U$, to obtain
\begin{equation}
E_{\text{sc}}=\text{Tr}[\mathcal{H}_{\text{sc}}(Q_{t},P_{t},q_{t}%
)]=H_{\text{sc}}(Q_{t},P_{t};q_{t}), \label{E_scl_final_v2}%
\end{equation}
where $H_{\text{sc}}$ is the semiclassical Hamiltonian (\ref{eq:H_sc}) from
Section~\ref{subsec:H_map}. Note that with this choice of $Q_{0}$, the
right-hand side of the generalized commutation relation (\ref{eq:QP_Hermitian}%
) becomes $i\hbar I_{D}$, in direct analogy with $[\hat{q},\hat{p}]=\hat
{q}\otimes\hat{p}^{T}-\hat{p}\otimes\hat{q}^{T}=i\hbar I_{D}$, but differs
slightly from Hagedorn's convention\cite{Hagedorn:1980,Hagedorn:1998} of
$2iI_{D}$, which would also fail to eliminate the factor $\hbar/2$ in the
energy (\ref{E_scl_final_v1}). If the exact potential is replaced with the
single-Hessian potential $V_{\text{SH}}$ [Eq.~(\ref{eq:SHLHA})], the matrix
function $\mathcal{H}_{\text{sc}}$ from Eq.$~$(\ref{eq:calH_sc}) becomes
independent of $q_{t}$, and so does $H_{\text{sc}}$. As discussed in
Section~\ref{subsec:H_map}, in this setting $H_{\text{sc}}(Q_{t},P_{t})$ is a
constant of motion, and, so is the semiclassical energy, since, according to
Eq.~(\ref{E_scl_final_v2}), it is equal to the semiclassical Hamiltonian
$H_{\text{sc}}$. Finally, in agreement with the derivation presented in
Sec.~\ref{subsec:shtga}, the total energy is conserved because it is equal to
the mapping Hamiltonian $H_{\text{map}}(q_{t},p_{t},Q_{t},P_{t})$ of
Eq.{~(\ref{eq:H_map}).}

\bibliographystyle{aipnum4-1}
\bibliography{biblio43,additions_SingleHessian}

\end{document}